# Spectropolarimetric evidence for a kicked supermassive black hole in the Quasar E1821+643

ANDREW ROBINSON[1,2], STUART YOUNG[1], DAVID J. AXON[1,2,3], PREETI KHARB[1], AND JAMES E. SMITH[4]

[1]Department of Physics, Rochester Institute of Technology, 84 Lomb Memorial Drive, Rochester, New York 14623, USA

[2]Centre for Astrophysics Research, Science & Technology Research Institute, University of Hertfordshire, Hatfield AL10 9AB, UK

[3]School of Mathematical & Physical Sciences, University of Sussex, Falmer, Brighton, BN2 9BH, UK

[4]Department of Physics & Astronomy, The Open University, Walton Hall, MK7 6AA, UK

## ABSTRACT

We report spectropolarimetric observations of the quasar E1821+643 ($z$=0.297), which suggest that it may be an example of gravitational recoil due to anisotropic emission of gravitational waves following the merger of a supermassive black hole (SMBH) binary. In total flux, the broad Balmer lines are redshifted by $\approx$1000 km s$^{-1}$ relative to the narrow lines and have highly red asymmetric profiles, whereas in polarized flux the broad H$\alpha$ line exhibits a blueshift of similar magnitude and a strong blue asymmetry. We show that these observations are consistent with a scattering model in which the broad-line region has two components, moving with different bulk velocities away from the observer and towards a scattering region at rest in the host galaxy. If the high velocity system is identified as gas bound to the SMBH, this implies that the SMBH is itself moving with a velocity ~2100 km s$^{-1}$ relative to the host galaxy. We discuss some implications of the recoil hypothesis and also briefly consider whether our observations can be explained in terms of scattering of broad-line emission originating from the active component of an SMBH binary, or from an outflowing wind.

1. INTRODUCTION

The formation of SMBH binary systems in the centers of large galaxies seems a likely consequence of galaxy mergers (Begelman et al. 1980). If the binary eventually coalesces, a recoil ('kick') velocity is imparted to the merged SMBH as a result of anisotropic emission of gravitational radiation (e.g., Favata et al. 2004). Following a recent breakthrough in numerical relativity (Campanelli et al. 2006; Baker et al. 2006), it has been shown that recoil velocities can reach ~ a few 1000 km s$^{-1}$; Campanelli et al. 2007a,b; González et al. 2007; Tichy & Marronetti 2007). Such speeds would cause large displacements of the SMBH from the center of its host galaxy, or in extreme cases, eject it entirely, with profound astrophysical and cosmological consequences (e.g., Merritt et al. 2004; Volonteri et al. 2010).

If the progenitor binary has an accretion disk, some of the gas would be retained by the recoiling SMBH. Gravitational recoil might then be detected observationally, as a displaced quasar (Merritt et al. 2004; Madau & Quataert 2004; Loeb 2007), or via Doppler shifting of emission lines from the retained gas (Bonning et al. 2007). To date, three objects have been proposed as recoil candidates: the quasars SDSSJ092712.65+294344.0 (Komossa et al. 2008) and SDSS J105041.35+345631.3 (Shields et al. 2009) which exhibit large relative Doppler shifts; and most recently, a peculiar double-nucleus galaxy CXOC J100043.1+020637 identified in the COSMOS survey (Civano et al. 2010). Here we present spectropolarimetric observations of the quasar E1821+643, which reveal another candidate, with the unique feature that Doppler shifts indicating a large relative velocity between the SMBH and its host galaxy are observed in both direct and scattered (polarized) light.

E1821+643 ($z$=0.297) is one of the most luminous quasars in the local universe, having an absolute visual magnitude $M_V$=–27.1 (Pravdo & Marshall 1984, Hutchings & Neff 1991). Its host galaxy is a large elliptical (Hutchings & Neff 1991; McLeod & McLeod 2001), of half-light radius $R_{1/2} \approx 14$ kpc and estimated total mass $M_{gal} \approx 2\times10^{12}$ M$_\odot$ (Floyd et al. 2004). Although the quasar is classified as radio-quiet, it has extended low surface brightness radio emission extending well beyond the host galaxy (Blundell & Rawlings 2001). On arcsecond scales, an apparent ~90° bend in the SW jet has been interpreted in terms of precession of the jet axis in an SMBH binary (Blundell et al. 1996; Blundell & Rawlings 2001), but large changes in jet direction may also result from a re-orientation of the SMBH spin axis following coalescence of the binary (Merritt & Ekers 2002).

We adopt the following cosmology: $H_0$= 73 km s$^{-1}$ Mpc$^{-1}$, $\Omega_M$ = 0.27 and $\Omega_\Lambda$ =0.73.

## 2. SPECTROPOLARIMETRY

The spectropolarimetry data were obtained using the ISIS dual-beam spectrograph mounted on the 4.2-m William Herschel Telescope on the nights of 1998 October 1 and 2, 2003 August 18 and 2004 August 12–15. Observing techniques and data reduction procedures are described by Smith et al. (2002), who also present a polarization spectrum obtained from the 1998 data. The spectrum displayed in Fig. 1 is the result of combining data from all 7 separate observations, after rebinning to a common spectral resolution of 3.4 Å. The total exposure time for the combined spectrum is 28000 s in the red ($\lambda > 6500$ Å), 8800 s in the blue ($\lambda < 6500$ Å).

The total flux spectrum is dominated by the broad emission lines of the Hydrogen Balmer series and the "big blue bump", usually attributed to the accretion disk. Also present are the narrow components of the Balmer lines and various forbidden lines typical of AGN, most prominently [OIII]$\lambda\lambda$ 4959, 5007. The narrow lines form a single redshift system, at an average redshift $z=0.2970\pm0.0003$, which we take as the rest frame of the host galaxy. The broad Balmer lines show two notable characteristics: firstly, the line peaks are redshifted by $\approx$1000 km s$^{-1}$ relative to the corresponding narrow lines (e.g., 1280 km s$^{-1}$ for broad H$\beta$); secondly, they have extremely asymmetric profiles, with red wings extending to Doppler velocities of at least 15400 km s$^{-1}$ relative to the rest frame wavelength. Landt et al. (2008) reported similar redshifts, of 1000–2000 km s$^{-1}$ for the broad components of H$\alpha$, H$\beta$ and several Paschen series lines.

E1821+643 is only weakly polarized, with an average degree of polarization of 0.21$\pm$0.03% at a position angle 140$\pm$5°. The average polarization position angle is approximately perpendicular to the arcsecond-scale radio source, which is oriented roughly NE–SW (PA ~ 20–40°). In detail, large variations in both the degree, $p(\lambda)$ and position angle, $\theta(\lambda)$, of polarization are clearly evident across the broad H$\alpha$ line in the combined spectrum. Such spectral variations indicate that the dominant polarization mechanism is scattering, most likely by free electrons, rather than dichroic absorption by interstellar dust (Smith et al. 2002). The most striking feature of the data is that, in polarized flux ($p(\lambda)\times F(\lambda)$), the broad H$\alpha$ line is both blueshifted relative to the host galaxy rest frame, and displays a strong blue asymmetry; the velocity shift and profile asymmetry are comparable in magnitude but *opposite in direction* to the shift and asymmetry seen in total flux (Fig. 2a). Although they cannot be discerned in the data, it seems reasonable to suppose that similar polarization structures are associated with the weaker lines of the Balmer series, or HeI λ5876, which have line profiles similar to that of Hα in total flux.

## 3. THE NATURE OF E1821+643

### 3.1 *Origin of the Doppler-shifted broad-lines*

In general, the Doppler shifting of a spectral feature in scattered light implies relative motion between the emission source and the scattering medium. A given scattered photon is subject to successive Doppler shifts, the first due to the relative velocity between the source and the scattering particle, the second due to the motion of the scatterer relative to Earth. A net blueshift of the scattered line profile could, by itself, be explained if the scattering region is undergoing infall towards the SMBH (and thus the broad-line emitting region, hereafter BLR), or if the BLR gas itself is flowing out towards the scattering region. In the case of E1821+643, however, it is also necessary to explain the reversal in the sense of the profile asymmetry (Fig. 2a). The simplest explanation of the latter is that the BLR contains regions of emitting gas moving with different bulk velocities relative to the scattering region.

The total flux Hα profile in E1821+643 can be modelled by two Gaussian components (Fig. 2b). The first (hereafter R1) is identified with the line peak and has a line-of-sight velocity 470±15 km s$^{-1}$ and velocity dispersion (FWHM) 3620±50 kms$^{-1}$. The second component (R2), which emits the extended red wing, has a higher line-of-sight velocity, 2070±50 km s$^{-1}$, and a larger FWHM, 7790±70 km s$^{-1}$. Both components are redshifted and hence moving away from the observer. The opposite asymmetry will be obtained in polarized flux if they are also moving *towards* the scattering medium. Then, as R2 has the largest redshift in total flux it will also have the largest blueshift in scattered flux, thus producing both the net blueshift and the reversed asymmetry as observed in polarized light.

### 3.2 Scattering model

The simplest scattering geometry consistent with the observations is illustrated in Fig. 3. The two BLR components move towards a stationary scattering region represented by a cone segment, with velocity vectors directed along its symmetry axis. Simulated polarization spectra were computed for this geometry using the scattering model described by Young (2000). The inclination of the line of sight to the symmetry axis, $\alpha$, and hence the bulk velocities of R1 and R2, can be constrained by matching the observed degree of polarization, while maintaining the bulk Doppler shifts of the two components in the total flux spectrum. We find that models with $110° \leq \alpha \leq 170°$ reproduce the main features of the data reasonably well (Fig. 4). The corresponding ranges in the velocities of the two BLR components are $480 \leq |v_{R1}| < 1370$ km s$^{-1}$ and $2100 \leq |v_{R2}| < 6050$ km s$^{-1}$, respectively. Essentially identical results are obtained for a bi-conical scattering region, as long as the BLR components are closer to the far-side than the near-side scattering cone. We also note that models in which R1 is stationary at the center of symmetry produce results consistent with the polarization spectra, although the fit to the total flux H$\alpha$ profile is significantly worse in this case.

The large change ($\Delta\theta \approx 25°$) in polarization position angle coincident with the red wing of the H$\alpha$ profile indicates a secondary source of polarization, which becomes prominent in this wavelength range because the scattered H$\alpha$ emission is Doppler-shifted to shorter wavelengths. Possible secondary polarization sources include dichroic absorption by dust along the line of sight to the quasar, or as seen in some active galactic nuclei, scattering by electrons situated close to the BLR, near the equatorial plane of the system (Smith et al. 2005). An example of the effect of including a compact equatorial scattering region in the model is shown in Fig. 4. However, whatever its origin, the presence of this secondary component of polarized light does not alter our main conclusions.

## 4. DISCUSSION

Our interpretation of the Hα polarization suggests that the high velocity dispersion component (R2) of the BLR in E1821+643 is undergoing bulk motion at a speed ~ few $\times 10^3$ km s$^{-1}$, relative to the host galaxy. We now consider the possibility that this may be a signature of gravitational recoil and also briefly outline two alternative scenarios.

### *4.1 Gravitational Recoil*

A "kicked" SMBH retains gas whose Keplerian velocity exceeds the recoil velocity (Merritt et al. 2006; Loeb 2007). If we identify Hα component R2 with BLR gas that has remained bound to the SMBH, its inferred bulk velocity, $v_{R2}$, gives the velocity of the SMBH at the current point in its post-kick trajectory. The scattering models are then consistent with recoil velocities ranging from 2100 km s$^{-1}$ up to the extrapolated maximum of 4000 km s$^{-1}$ deduced from numerical relativity experiments (Campanelli et al. 2007b; this corresponding to a viewing angle α ≈ 120°). If it fails to escape the host galaxy, the recoiling SMBH will oscillate about the center of mass and for a large kick, relatively little energy is lost during the first few oscillations (Gualandris & Merritt 2007). The high velocity inferred suggests that we are seeing the SMBH either soon after coalescence, in its initial outward trajectory, or when it is approaching the galaxy core on the return trajectory of the first or second oscillation, at a velocity approaching that of the initial kick.

The former scenario is favored if we associate the ~ 1 arcsec scale radio source (Blundell et al. 1996) with a re-oriented post-coalescence jet. The jet lifetime then gives an estimate of the time elapsed since recoil: $t_{jet} \sim 7000 \ yr/(\beta_{jet} \sin i)$, where $\beta_{jet}$ is the jet advance speed in units of the speed of light, $i$ is the inclination of the jet axis to the line of sight and we have taken the projected length of the jet to be 0.5" (≈ 2 kpc). As $\beta_{jet} \sin i \leq 1$, the distance travelled by the recoiling SMBH must be $d \geq 18$ pc. In fact, jet advance speeds are likely to be $\ll c$; for example, Scheuer (1995) finds that typically $\beta_{jet} < 0.1c$, even for powerful radio sources. This would imply $t_{jet} \geq 10^5$ yr and $d \geq 180$ pc, for E1821+643.

It is reasonable to assume that the arcsecond-scale radio source traces the spin axis of the SMBH on the sky plane. The roughly perpendicular orientation of the plane of polarization with respect to the jet direction then implies that the spin angular momentum is aligned to within $\approx 20$–$30°$ of the recoil velocity vector. This is consistent with statistical studies of the coalescence of spinning black holes (Lousto et al. 2010), which indicate firstly, that the recoil velocity is preferentially aligned (or counter-aligned) with the orbital angular momentum of the original binary and secondly, that the spin of the recoiling SMBH tends to be slightly misaligned by $\approx 25°$ with the orbital angular momentum (for an equal mass progenitor binary; the probability distribution is broader for high mass ratios).

If the recoil velocity is indeed aligned with the binary angular momentum axis, the scattering region must be located above the plane of the original binary. As the polarized (scattered) flux has the same overall shape as the total flux spectrum (Fig. 1), the scattering particles are most likely electrons. There is evidently no lack of ionized gas present in the galaxy; E1821+643 resides at the center of a strong cooling flow (Russell et al. 2010) and also has a large extended emission line region (Fried 1998).

While it is natural to identify R2 with that part of the BLR that remains bound to the SMBH, the nature of the low velocity component, R1, is less clear. The velocity dispersion of this component is $\sim v_{R2}$, as would be expected for the gas left behind by the impulsive ejection of the SMBH. Thus, it may represent the marginally bound or unbound portion of the BLR, which has been left in the "wake" of the SMBH (c.f., Shields & Bonning 2008), or the outer regions of a possibly much larger circumbinary disk. However, our Gaussian decomposition indicates that this component contributes ≈40% of the broad Hα flux. If both components are photoionized by the displaced quasar (i.e., the recoiling SMBH and its remnant accretion disk), they must have similar covering fractions as seen from the quasar. This may be hard to reconcile with a displacement greater than a few 10's of parsecs, suggesting that only a few $\times 10^4$ years have elapsed since recoil.

A reasonable objection to the recoil hypothesis is that it is statistically unlikely. Large recoil velocities require specific pre-coalescence binary configurations (mass ratio ≈ 1, large spin components in the orbital plane; Campanelli et al. 2007b), which are expected to occur infrequently even in dry mergers (Lousto et al. 2010; Volonteri et al. 2010), and may even be physically inhibited in gas-rich systems (Bogdanovich et al. 2007; Dotti et al. 2010). Moreover, as discussed above, we must also be observing the SMBH relatively soon after coalescence.

### 4.2  *Binary black hole*

Bogdanović et al. (2009) and Dotti et al. (2009) have argued that the recoil candidate SDSSJ092712.65+294344.0 (Komossa et al. 2008) can be explained as an unequal mass black hole binary system, in which only the secondary is active. The broad emission lines come from gas bound to the secondary SMBH, while the narrow lines are predominately emitted by a circum-binary disk, which also supplies the accretion flow to the secondary. In applying this scenario to E1821+643, the bulk redshift of the broad Balmer lines can be attributed to orbital motion while relativistic effects might explain the asymmetric shapes of the line profiles in direct light if the emission comes from the accretion disk itself (e.g., Chen et al. 1989). Scattering in the accretion flow from the circumbinary disk would tend to produce a net blueshift in polarized light, as required by the observations. However, without detailed modelling, it is not clear if the blue asymmetry seen in polarized flux can be reproduced in such a scattering geometry. Furthermore, given that the radio jet is aligned with the spin axis of the active secondary, and that this is perpendicular to the orbital plane, this scattering geometry would result in polarization aligned with the radio jet direction, contrary to what is observed.

*4.3     Outflowing BLR*

Another possible explanation for E1821+ 643 is that the broad Hα lines are emitted in an outflowing wind. The red-shift and asymmetry of the line profiles in total flux can, in principle, be explained in terms of a bi-polar wind if the emitting clouds are individually optically thick to their line radiation (e.g., Capriotti 1981); the observed emission comes preferentially from the illuminated faces of clouds in the receding side of the flow. However, a *surrounding* scattering region will likewise "see" only emission from receding clouds on the opposite side of the flow, producing redshifted, rather than blueshifted, profiles in scattered flux. An intrinsically one-sided, jet-like flow having "slow" and "fast" components corresponding to R1 and R2, respectively, would be exactly analogous to the model outlined in Section 3 and would therefore produce a similar polarization spectrum. However, there is no clear physical justification for a uni-directional wind. The structure of the radio source argues against this scenario; there is clear evidence of sustained outflow of synchrotron-emitting plasma in anti-parallel directions at both sub-arcsecond and ~30" scales (Blundell & Rawlings 2001). Nevertheless, we cannot exclude the possibility that other configurations of emitting and scattering flows may be capable of reproducing the observations.

5.     CONCLUSIONS

The optical spectrum of the luminous quasar E1821+643 is unusual in that the broad emission lines are both redshifted and strongly red-asymmetric. Even more remarkably, for Hα at least, the shift and asymmetry are reversed in polarized light. These observations can be understood if a major component of the BLR is moving with a velocity ~2100 km s$^{-1}$ with respect to a scattering medium at rest in the host galaxy. We propose that this may be a signature of gravitational recoil following coalescence of an SMBH binary system, the high velocity gas representing that portion of the BLR that has remained bound to the recoiling, merged SMBH. This interpretation, however, demands that we are observing the SMBH at an early stage in its trajectory following a high velocity kick, circumstances that appear statistically unlikely to be found in a relatively nearby object. It also seems unlikely that the recoil hypothesis can be confirmed by direct measurement of the angular displacement of the quasar nucleus; the linear displacement is probably small (< 1 kpc) and the recoil velocity vector is closely aligned with the line of sight. As noted by Blecha & Loeb (2008) these are general problems for identification of gravitational recoil candidates through large doppler shifts.

We have also considered whether the observations might be explained by models in which the BLR is associated with the secondary in an SMBH binary system, or is part of a wind associated with a single SMBH at rest at the center of the host galaxy. It is not clear, without more detailed modelling, if a scattering geometry exists that can fully account for observed polarization spectrum in either case. Nevertheless, since observational verification of high velocity gravitational recoils is of great interest, both as confirmation of the numerical relativity results and as signposts of black hole coalescence, these alternative scenarios merit more detailed investigation.

This material is based upon work supported in part by the National Science Foundation under Award No. AST-087633. The William Herschel Telescope is operated on the island of La Palma by the Isaac Newton Group in the Spanish Observatorio del Roque de los Muchachos of the

Instituto de Astrofísica de Canarias. We thank the anonymous referee for insightful comments, which helped to improve this paper.

REFERENCES

Baker, J. G., Centrella, J., Choi, D.-I., Koppitz, M., & van Meter, J. R. 2006, Phys. Rev. Lett., 96, 111102.

Begelman, M. C., Blandford, R. D. & Rees, M. J. 1980, Nature 287, 307.

Blecha, L. & Loeb, A. 2008, MNRAS, 390, 1311

Blundell, K. M., Beasley, A. J., Lacy, M. & Garrington, S.T. 1996, ApJ 468, L91.

Blundell, K.M. & Rawlings, S. 2001, ApJ 562, L5.

Bogdanović, T., Eracleous & M. Sigurdsson, S. 2009, ApJ 697, 288.

Bogdanović, T., Reynolds, C. S. & Miller, M. C. 2007, ApJ 661, 147.

Bonning, E. W., Shields, G. A. & Salviander, S. 2007, ApJ 666, L13.

Campanelli, M., Lousto, C., Marronetti, P. & Zlochower, Y. 2006, Phys. Rev. Lett. 96, 111101.

Campanelli, M., Lousto, C., Zlochower, Y. & Merritt, D. 2007a, ApJ, 659, L5.

Campanelli, M., Lousto, C. O., Zlochower, Y., & Merritt, D. 2007b, Phys. Rev. Lett., 98, 231102.

Capriotti, E., Foltz, C. & Byard, P. 1981, ApJ, 245, 396

Chen, K., Halpern, J. P. & Filippenko, A. V. 1989, ApJ 339, 742.

Civano, F., et al., 2010, ApJ, submitted (http://arxiv.org/abs/1003.0020).

Dotti, M., Montuori, C., Decarli, R., Volonteri, M., Colpi,M. & Haardt F. 2009, MNRAS 398, L73.

Dotti, M.; Volonteri, M.; Perego, A.; Colpi, M.; Ruszkowski, M.; Haardt, F. 2010, MNRAS, 402, 682

Favata, M., Hughes, S. A. & Holz, D. E. 2004, ApJ 607, L5.

Floyd, D. J. E., Kukula, M. J., Dunlop, J. S., McLure, R. J., Miller, L., Percival, W. J., Baum, S. A. & O'Dea, C. P. 2004, MNRAS 355, 196.

Fried, J.W. 1998, A&A, 331, L73

González, J. A., Hannam, M., Sperhake, U., Brügmann, B., & Husa, S. 2007, Phys. Rev. Lett., 98, 231101

Gualandris, A. & Merritt, D. 2007, ApJ 678, 780.

Hutchings, J.B. & Neff S.G. 1991, AJ 101, 2001.

Komossa, S., Zhou, H. & Lu, H. 2008 ApJ, 678, L81.

Landt, H., Bentz, M. C., Ward, M. J., Elvis, M., Peterson, B. M., Korista, K. T., Karovska, M. 2008, ApJS, 174, 282.

Loeb, A. 2007, Phys. Rev. Lett. 99, 041103.

Lousto, C. O., Nakano, H., Zlochower, Y. & Campanelli M. 2009, Phys Rev D.81.084023

Madau, P., & Quataert, E. 2004, ApJ 606, L17.

McLeod, K. K. & McLeod, B.A. 2001, ApJ, 546, 782.

Merritt, D. & Ekers, R. D. 2002, Science 297, 1310.

Merritt, D., Milosavljević, M., Favata, M., Hughes, S. A. & Holz, D. E. 2004, ApJ 607, L9.

Merritt, D., Storchi-Bergmann, T., Robinson, A., Batcheldor, D., Axon, D.J. & Cid Fernandes, R. 2006, MNRAS 367, 1746.

Pravdo, S. H. & Marshall, F. E. 1984, ApJ 281, 570.

Russell, H. R, Fabian, A. C, Sanders, J. S, Johnstone, R. M, Blundell, K. M, Brandt, W. N, Crawford, C. S, 2010, MNRAS, 402, 1561

Scheuer, P. 1995, MNRAS, 277, 331

Shields, G. A. & Bonning, E. W. 2008, ApJ 682, 758.

Shields, G.A., et al., 2009, ApJ 707, 936.

Smith, J. E., Robinson, A., Young, S., Corbett, E. A., Giannuzzo, M. E., Axon, D. J. & Hough, J.H. 2002, MNRAS 335, 773.

Smith, J. E., Robinson, A., Young, S., Axon, D. J. & Corbett, E. A. 2005, MNRAS 359, 846.

Tichy, W. & Marronetti, p. 2007, Phys. Rev. D 76, 061502.

Volonteri, M., Gultekin, K., Dotti, M. 2010, MNRAS, in press (http://arxiv.org/abs/1001.1743).

Young, S. 2000, MNRAS 312, 567.

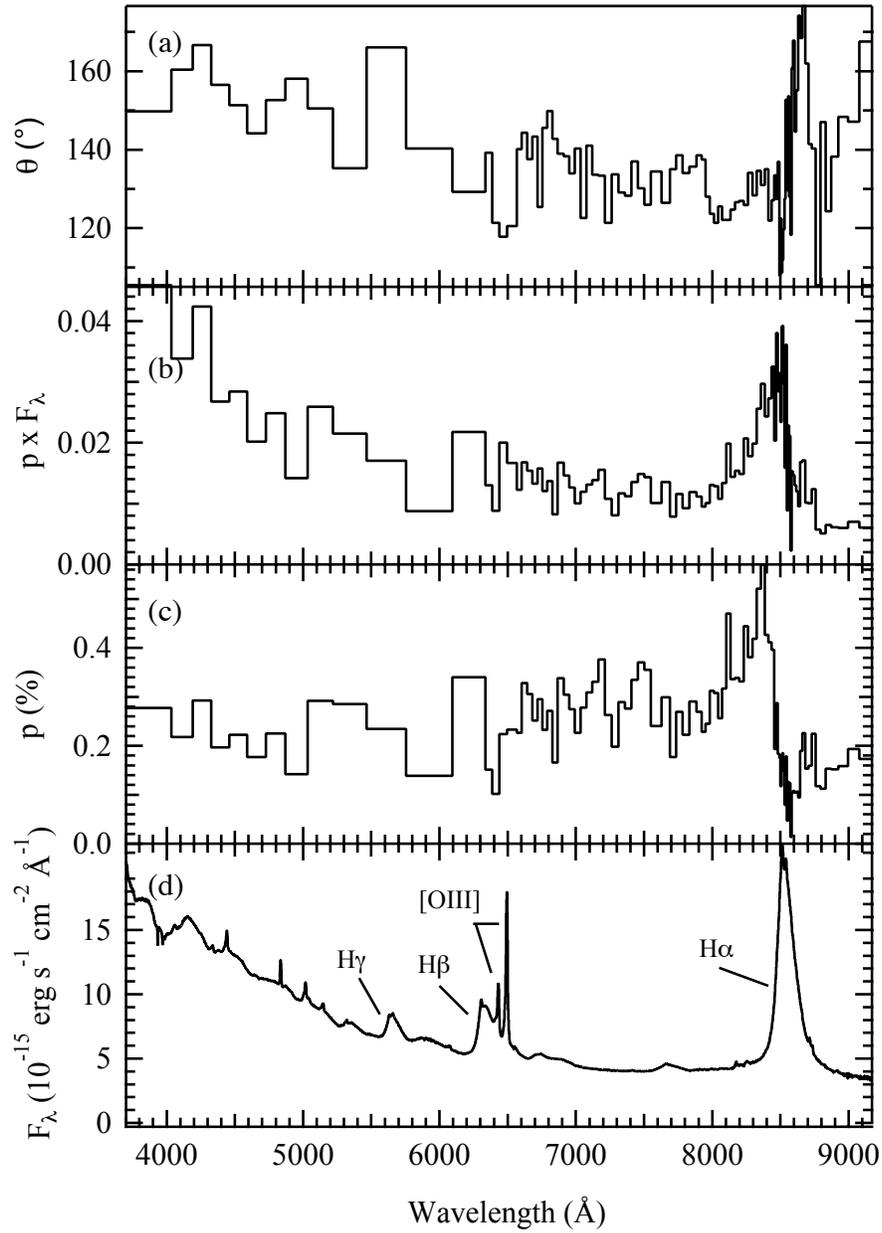

FIG 1. —The polarization spectrum of the Quasar E1821+643: (a) the position angle of polarization; (b) the polarized flux spectrum; (c) the degree of polarization; (d) the total flux spectrum. The polarized flux density has the same units as the total flux density. The polarization data have been re-sampled into bins with an error of 0.05% in degree of polarization.

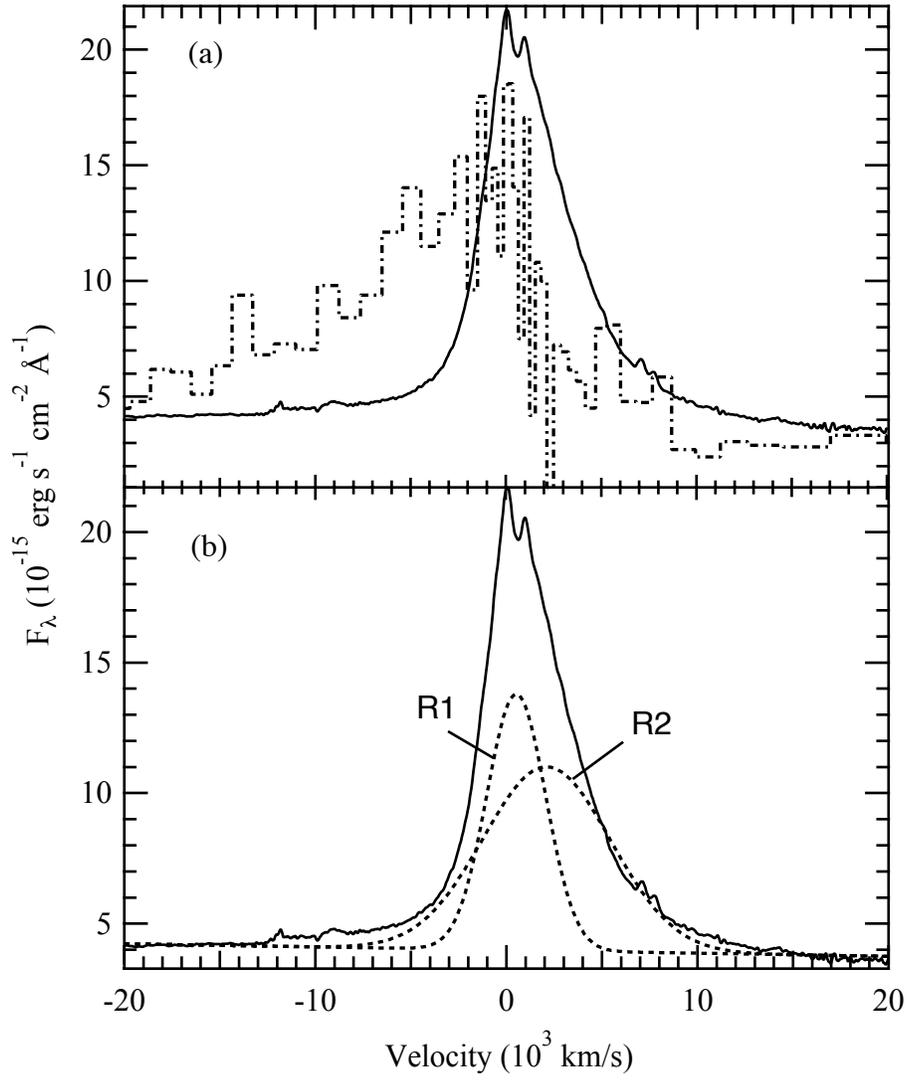

FIG 2. —The Hα line profile in velocity space. The velocity zero is defined at the average redshift of the narrow emission lines. (a) Comparison of the total (solid line) and polarized flux (dot-dashed line) profiles. The polarized flux spectrum has been by scaled by the inverse of the average polarization. (b) Total flux profile (solid) with the fitted Gaussian components R1 and R2 (dashed lines).

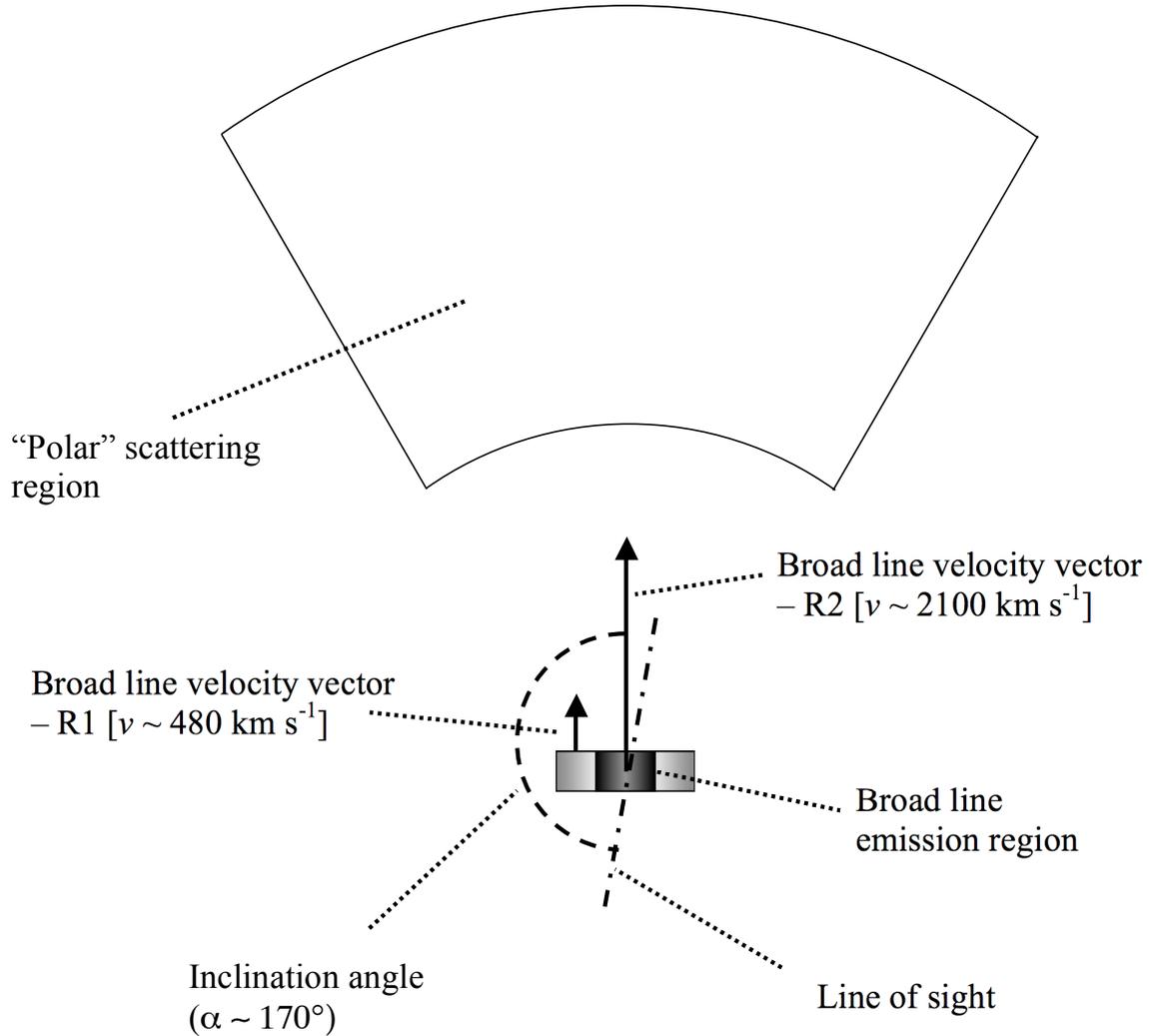

FIG 3.—A schematic illustration of the emission and scattering geometry invoked to explain the polarization properties of E1821+643. The broad-line emission originates in two components of gas characterised by different bulk velocities and velocity dispersions (R1 and R2). Both velocity vectors are aligned with the symmetry axis of the scattering region and the observer's line of sight is inclined at angle α to this axis. The scattering is far-field in the sense that the emission regions effectively constitute a single point source as viewed from the scattering region.

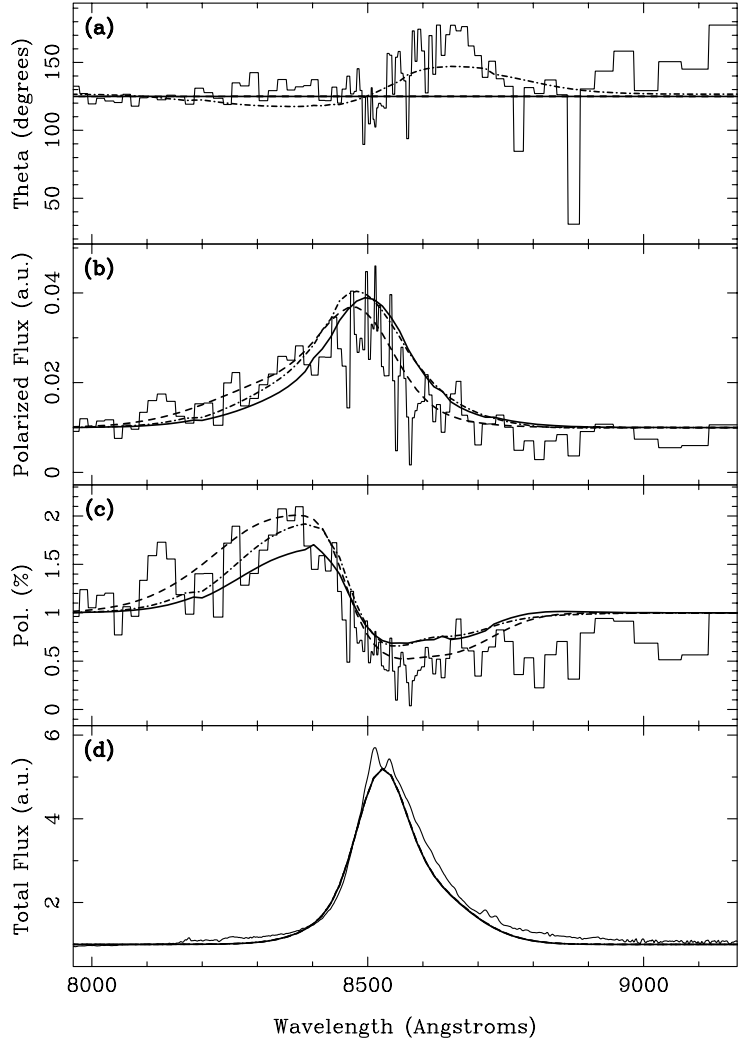

FIG 4. —Comparison of simulated and observed polarization spectra for the Hα region. The panels are the same as in Fig. 1, with the exception that the total and polarized flux densities are given in arbitrary units (a.u.), while the degree of polarization is normalized to 1%. Models for viewing angles ranging from α = 170° (solid heavy line) to α=110 (dashed line) are consistent with the data. The large change in polarization position angle coincident with the weakly polarized red wing of the Hα profile can be explained by including a second source of polarization. The effect of including a compact equatorial scattering region is shown for the α=170° model (dot-dashed line).